\documentstyle[prb,aps,preprint,eqsecnum,tighten]{revtex}
\begin{document}
\draft
\title{Dynamical correlation functions
of one-dimensional superconductors and
Peierls and Mott insulators
\footnote{Dedicated to Professor Johannes Zittartz
on the occasion of his 60$^{th}$ birthday}
}
\author{Johannes Voit}
\address{Theoretische Physik 1, Universit\"{a}t Bayreuth, 
D-95440 Bayreuth (Germany)}
\date{\today}
\maketitle
\begin{abstract}
I construct the spectral  function of the Luther-Emery model which
describes one-dimensional fermions with one gapless and one gapped 
degree of freedom, i.e. superconductors and Peierls and Mott insulators,
by using symmetries, relations to other models, and known limits. 
Depending on the relative magnitudes of the charge and spin 
velocities, and on whether a charge or a spin gap is present, I find
spectral functions differing in the number of singularities and presence
or absence of anomalous dimensions of fermion operators. I find, for a Peierls
system, one singularity with anomalous dimension and one finite
maximum; for a superconductor two singularities with anomalous dimensions;
and for a Mott insulator one or two singularities without anomalous 
dimension. In addition, there are strong shadow bands. I 
generalize the construction to arbitrary dynamical multi-particle
correlation functions. The main aspects of this work are in agreement
with numerical and Bethe Ansatz calculations by others. I also
discuss the application to photoemission experiments on 1D Mott insulators
and on the normal state of 1D Peierls systems, and propose the Luther-Emery
model as the generic description of 1D charge density wave systems with
important electronic correlations.
\end{abstract}

\narrowtext

\section{Motivation}
\label{moti}
Non-Fermi liquid behavior in correlated fermion systems is an exciting topic
of current research. One-dimensional (1D) correlated electrons 
(more precisely: one-dimensional quantum systems with gapless excitations)
are a paradigmatic example
of non-Fermi liquids: their low-energy excitations are not quasi-particles but
rather collective charge and spin density fluctuations which obey each to 
their proper dynamics \cite{myrev}. The key features of these 
``Luttinger liquids'' \cite{Haldane} are 
(i) anomalous dimensions of operators producing correlation
functions with non-universal power-laws, parametrized by one renormalized
coupling constant $K_{\nu}$ per degree of freedom $\nu=\rho$ (charge), $\sigma$
(spin) 
which have the status of the Landau parameters familiar from Fermi liquid
theory; (ii) charge-spin separation, leading to a fractionization 
of an electron into charged, spinless, and neutral, 
spin-carrying collective excitations, 
with different dynamics determined by velocities
$v_{\rho} \neq v_{\sigma}$. Each of these features 
leads to (iii) absence of fermionic quasi-particles. 
Responsible
are the electron-electron interaction which is marginal
in one dimension and therefore transfers nonvanishing momentum in scattering
processes at all energy scales, and the nesting properties of the 1D Fermi
surface. They produce divergent $2k_F$ charge and spin 
density fluctuations which then interfere with Cooper-type superconducting
fluctuations. 

All three features 
clearly show up in the single-particle spectral function 
\cite{specll,sanseb,sm}
\begin{equation}
\label{specdef}
\rho(q, \omega) = - \pi^{-1} {\rm Im} \: G(k_F+q, \mu + \omega)
\end{equation}
which can be measured (within the ``sudden approximation'')
by angle-resolved photoemission (ARPES) [with bad
angular resolution, one essentially measures $
N(\omega) = \sum_q \rho(q,\omega)$ and is able to probe only features 
(i) and (iii)]. The spectral function 
is purely incoherent \cite{specll,sanseb,sm}, at best
with peaks at the dispersion energies of the elementary charge and spin
excitations, indicating that the electron behaves as a composite particle
built on more elementary excitations.
In Eq.~(\ref{specdef}), 
$G$ is the Fourier transform of the retarded electronic Green's function
\begin{equation}
\label{grefct}
G(xt) = -i \Theta(t) \langle \{ \Psi(xt), \Psi^{\dagger}(00) \} \rangle \;\;,
\end{equation} 
$k_F$ the Fermi wave number, and $\mu \; (=0)$ is the chemical potential. 

Much experimental effort has been devoted to studying and attempting to
``prove'' Luttinger liquid correlations in various quasi-1D systems.
Examples are organic conductors of the family based on the molecule
$TMTSF$ (Bechgaard salts) where both
NMR \cite{claude} and (partially) photoemission \cite{tm} have provided evidence
in favor of a Luttinger liquid picture, quantum wires fabricated into
semiconductor nanostructures \cite{quawi}, or edge states in the fractional
quantum Hall effect \cite{edge}. In all cases however, there appear
to be problems
with the precise values of the parameter $K_{\rho}$ derived, or with 
some other aspects of the interpretation in terms of a Luttinger liquid.
It is not clear to date to what extent these discrepancies are due to
the neglect of some experimentally important factor in the theory (such
as, e.g. three-dimensionality or electron-phonon coupling in the chain
systems, or deviations from the special filling factors in the quantum
Hall edge states), or indicative of more fundamental problems either with
theory or experiment.

1D (organic and inorganic) 
charge density wave (CDW) systems apparently could provide an alternative
field of search for these typically one-dimensional correlations. 
Photoemission indeed has
produced results \cite{dardel} similar to the Bechgaard salts when performed
with low angular resolution. With high angular resolution, a broad 
dispersing feature has been identified in $(TaSe_4)_2I$ \cite{tase}
while \em two \rm such signals have been measured in the blue bronze
$K_{0.3} Mo O_3$ \cite{gweon}. Even though the actual situation in 
$K_{0.3} Mo O_3$ may be slightly more complictated because there are
two almost degenerate bands cutting the Fermi energy, it is clearly 
of importance to first understand the photoemission spectrum expected
from the metallic phase of a single-band CDW material. Finally, while
this paper was prepared, new experiments on the organic two-chain
conductor $TTF-TCNQ$ became available which clearly show dispersing 
signals both on the $TTF$ and $TCNQ$ chains with very unusual lineshapes 
\cite{ttf}.
Specifically, the $TCNQ$ signals are somewhat similar to $K_{0.3} Mo O_3$,
and we know from independent experiments that there are strong $2k_F$-CDW
fluctuations on this chain in the metallic state \cite{ttfcdw}. (The
$TTF$-chain exhibits strong $4k_F$-CDW fluctuations at very high temperature
and is expected to be a Luttinger liquid.) 

The association of the two dispersing signals of 
$K_{0.3} Mo O_3$ with the charge and spin
excitations of a Luttinger liquid is suggestive. As I will explain in the
next section in more detail, it is incompatible, however, with the CDW
transitions observed in these materials. This incompatibility 
motivates the consideration
of the Luther-Emery model and is at the origin of the work reported here. 
Section II will discuss this model, its
generic role as a low-energy fixed-point of 1D quantum systems which have
both gapped and gapless degrees of freedom, and the picture we had of its
correlations prior to this work. 

Recently, photoemission experiments also have been performed on the
1D Mott insulator $Sr Cu O_2$ \cite{kim}. In Mott insulators, the charge
fluctuations are gapped while the spins remain gapless. Their low-energy
physics, therefore, can again be described by a Luther-Emery model, and
our theory can be adapted to study the spectral functions of 1D Mott 
insulators. Earlier, angle-integrated photoemission on $BaVS_3$ has been
interpreted as evidence for a Luttinger liquid \cite{fuji}. The behavior
of the conductivity, however, is more insulator-like, and present theory
might be of interest there, too.

Section III presents the construction of the single-particle spectral
function (\ref{specdef}). In  
Section \ref{specfle} I present results for the spectral functions of the
spin-gapped Luther-Emery model, i.e. 1D Peierls systems and superconductors.
In Section \ref{specfmott}, the spectral functions of 1D Mott insulators
are presented. Section \ref{other} shows how the construction procedure
of Section \ref{construction} can be generalized to arbitrary correlation
functions of local operators. I compare my results with information from
other studies in Section \ref{colleagues} and use them for an interpretation
of published experiments in Section \ref{appl}. I conclude with a short summary
and a brief perspective.
Partial results have
been presented earlier \cite{brief1,ucla}. 

\section{The Luther-Emery model}
\label{lemodel}
The Luther-Emery model extends the Luttinger model by including the
backscattering of electrons across the Fermi surface. Its Hamiltonian
is \cite{luthem}
\begin{eqnarray}
\label{hlutt}
H & = & H_0 + H_1 + H_2 + H_4 \;\;\;, \\
\label{hfree}
H_0 & = & \sum_{r,k,s} v_F (rk - k_F) 
: c^{\dag}_{rks} c_{rks} : \;\;\;, \\
\label{h1}
H_{1} & = & \sum_{s,s'} 
\left[ g_{1\|} \delta_{s,s'} + g_{1\perp} \delta_{s,-s'} \right]
\int_0^L \! dx : \Psi_{+,s}^{\dag}(x)
\Psi_{-,s}(x) \Psi_{-,s'}^{\dag}(x) \Psi_{+,s'}(x) : \\
\label{h2}
H_2 & = & \frac{1}{L} \sum_{p,s,s'} \left[ g_{2\|} \delta_{s,s'} +
g_{2\perp} \delta_{s,-s'} \right] \rho_{+,s}(p) \rho_{-,s'}(-p) \;\;\;, \\
\label{h4}
H_4 & = & \frac{1}{2L} \sum_{r,p,s,s'} \left[ g_{4\|} \delta_{s,s'} +
g_{4\perp} \delta_{s,-s'} \right] : \rho_{r,s}(p) \rho_{r,s'}(-p) : 
\;\;\;. 
\end{eqnarray}
$c_{rks}$ describes fermions with momentum $k$ and spin $s$ on the
two branches ($r=\pm$) of the dispersion varying linearly [$\varepsilon_r(k)
= v_F (rk - k_F)$] about the two Fermi points $\pm k_F$, 
$\Psi_{r,s}(x)$ is its Fourier transform, and 
\begin{equation}
\label{rhoeq}
\rho_{r,s}(p) 
=  \sum_k : c^{\dag}_{r,k+p,s} c_{r,k,s} : 
=  \sum_k \left( c^{\dag}_{r,k+p,s} c_{r,k,s} -
\delta_{q,0} \langle c^{\dag}_{r,k,s} c_{r,k,s} \rangle_0 \right) 
\end{equation}
is the density fluctuation operator which obeys a bosonic algebra
\begin{equation}
\label{comm}
[\rho_{r,s}(p), \rho_{r',s'}(-p') ]  = - \delta_{r,r'} \delta_{s,s'}
\delta_{p,p'} \frac{rpL}{2 \pi} \;\;\;.
\end{equation}
The Luttinger model is obtained for $g_1=0$ and includes only
forward scattering. 

In one dimension, fermions can be transformed into bosons, and for the
Luttinger model, there is an exact operator identity relating a fermion
operator $\Psi_{rs}(x)$ to the bosonic density fluctuations (\ref{rhoeq})
\cite{myrev,Haldane}. For our purposes, the approximate expression
\begin{equation}
\label{bos}
\Psi_{rs}(x) \sim \lim_{\alpha \rightarrow 0}
\frac{ e^{irk_Fx} }{ \sqrt{2 \pi \alpha} } \exp \left( \frac{-i}{\sqrt{2}} 
\left[ r \Phi_{\rho}(x) - \Theta_{\rho}(x)
+ s \left\{ r \Phi_{\sigma}(x) - \Theta_{\sigma}(x) \right\} 
\right] \right) \;\;\;.
\end{equation}
with the two phase fields 
\begin{equation}
\label{phi}
\Phi_{\nu}(x) = - \frac{i\pi}{L} \sum_{p \neq 0 } 
\frac{e^{-\alpha \mid p \mid / 2 -ipx}}{p} \left[\nu_+(p)+\nu_-(p) \right]
\;\;\;,
\end{equation}
and
\begin{equation}
\label{theta}
\Theta_{\nu}(x) = \frac{i\pi}{L} \sum_{p \neq 0 } \frac{e^{-\alpha
\mid p \mid / 2 -ipx}}{p} \left[\nu_+(p) - \nu_-(p) \right] \;\;\;,
\end{equation}
found earlier by Luther and Peschel \cite{lupe}, is sufficient. 

This formula allows for a boson representation of the Hamiltonian and
of all correlation functions. Before, it is important, however, to 
recall the physics of the phase fields
$\Phi_{\nu}(x)$ and $\Theta_{\nu}(x)$ in (\ref{bos})
\cite{myrev,ijmb,halfl}. The charge density 
\em fluctuation \rm operator is related to $\Phi_{\rho}(x)$ by
$\sum_r \rho_r(x) = - \pi^{-1} \partial \Phi_{\rho}(x) / \partial x$, 
and likewise for spin $\sigma$. When
an additional particle is inserted into the system, a kink of amplitude
$\pi$ is formed in $\Phi_{\nu}(x)$. These fields therefore 
describe the scattering phase shifts of the particles present in the
system, generated by the particles added. The operators inserting 
the particles are exponentials of the dual fields $\Theta_{\nu}(x) = \int 
\Pi_{\nu}(x) dx$, where $\Pi_{\nu}$ is the momentum 
conjugate to $\Phi_{\nu}$: $[\Pi_{\nu}(x) , \Phi_{\nu}(x') ] =-i \delta(x-x')$.
In a general fluctuation operator
whose correlation function we wish to evaluate,
the prefactor of $i\Theta_{\nu}/\sqrt{2}$ measures the number of 
$\nu$-particles it inserts into the system while the prefactor of
$i\Phi_{\nu}/\sqrt{2}$ 
measures the number of $\nu$-particles it rearranges at constant
total $\nu$-particle number to generate the desired fluctuation. 
By $\nu$-particle, we label, in the first place, 
\begin{equation}
\label{psinu}
\Psi_{r\nu}(x) = (2 \pi \alpha)^{-1/4} \exp\{-i[r \Phi_{\nu}(x)
- \Theta_{\nu}(x)]/\sqrt{2}\} \;\;\;,
\end{equation}
the slowly-varying charge or spin part of the fermion operators $\Psi_{rx}(x)$
but, with phase factors reflecting the appropriate Fermi seas, these particles
will describe the holons and the spinons of the 1D Bethe-Ansatz soluble models.

The boson form of the Luther-Emery Hamiltonian becomes
\begin{eqnarray}
\label{hnu}
H_0 + H_4 & = & \frac{1}{L} \sum_{\nu r p \neq 0} 
(\pi v_F + g_{4 \nu}) : \nu_r(p) \nu_r(-p) : \\
H_{1,\|} + 
H_2 & = & \frac{1}{L} \sum_{\nu p} (2g_{2\nu} -g_{1\|})
\nu_+(p) \nu_-(-p) \;\;\;, \\
\label{h1pbos}
H_{1\perp} & = & \frac{2 g_{1\perp}}{(2 \pi \alpha)^2} \int \! dx
\cos \left[ \sqrt{8} \Phi_{\sigma}(x) \right] 
\;\;\; .
\end{eqnarray}
$\nu_r(p)$ are the operators for the charge and spin densities
\begin{eqnarray}
\rho_{r}(p) & = & \frac{1}{\sqrt{2}} \left[\rho_{r,\uparrow}(p) +
\rho_{r,\downarrow}(p) \right] \;\;\;, \nonumber \\
\label{chsdens}
\sigma_{r}(p) & = & \frac{1}{\sqrt{2}} \left[\rho_{r,\uparrow}(p) -
\rho_{r,\downarrow}(p) \right] \;\;\; ,
\end{eqnarray}
and the interactions have been transformed as
\begin{equation}
g_{i\rho} = \frac{1}{2} \left( g_{i\|} + g_{i\perp} \right) \;\;\; , \;\;\;
g_{i\sigma} = \frac{1}{2} \left( g_{i\|} - g_{i\perp} \right) \;\;\; .
\end{equation}
Diagonalizing the Luttinger part (i.e. $H$ excluding $H_{1\perp}$)
generates the renormalized velocities of the collective charge and
spin excitations and their stiffness constants
\begin{equation}
\label{knu}
v_{\nu} = \sqrt{\left[ v_F + \frac{g_{4\nu}}{\pi} \right]^2
- \left[\frac{g_{2\nu}-g_{1\|}/2}{\pi}\right]^2} \;, \;\;\;
K_{\nu} = \sqrt{\frac{\pi v_F + g_{4\nu}-g_{2\nu}+g_{1\|}/2}{\pi 
v_F+g_{4\nu}+g_{2\nu}-g_{1\|}/2}} \;\;\;.
\end{equation}
The phase fields transform as
$\Phi_{\nu}(x)  \rightarrow  \Phi_{\nu}(x) \sqrt{K_{\nu}}$ and 
$\Theta_{\nu}(x) \rightarrow  \Theta_{\nu}(x) / \sqrt{K_{\nu}}$. The main
effect of the $g_4$-interaction is a renormalization of $v_{\nu}$. We therefore
drop $H_4$ from explicit consideration in the following, and always assume
correctly renormalized velocities $v_{\nu}$.

For $K_{\sigma}-1$ sufficiently large with respect to $|g_{1\perp}|$, 
backscattering is irrelevant, and the Luther-Emery model reduces to 
a Luttinger liquid. Its renormalized value of $K_{\sigma}$ can be
calculated, e.g. by perturbative renormalization group \cite{chui}
which is well-controlled in this case or, if applicable, fixed to unity
by the requirement of spin-rotation invariance. Charge and 
spin excitations are gapless, and 
depending on the value of $K_{\rho}$, the dominant are spin density wave
(SDW, $K_{\rho}<1$, repulsive forward scattering) or triplet pairing
(TS, $K_{\rho} > 1$, attractive forward scattering). Charge density wave (CDW)
and singlet superconducting (SS) fluctuations, respectively, are subdominant.

The backscattering Hamiltonian $H_{1\perp}$ is, for $K_{\sigma}-1$
small enough compared to $|g_{1\perp}|$, a
relevant perturbation and opens a gap $\Delta_{\sigma}$ in the spin excitation
spectrum 
\begin{equation}
\label{esigma}
\varepsilon_{\sigma}(q) = \pm \sqrt{v_{\sigma}^2 q^2 + \Delta_{\sigma}^2}
\;\;\;.
\end{equation}
Luther and Emery have
shown that for the special value $K_{\sigma}=1/2$, the interaction Hamiltonian
$H_{1\perp}$
(\ref{h1pbos}) can be represented as a bilinear in spinless fermions, 
using the bosonization formula (\ref{bos}) for spinless fermions
(multiply the argument of the exponential by $\sqrt{2}$ and drop the
$\sigma$-fields), and
diagonalized \cite{luthem}. On this Luther-Emery line $K_{\sigma}=1/2$,
the gap is computed exactly to be $\Delta_{\sigma} = |g_{1\perp}| / 2 \pi
\alpha$ [$\alpha$ is an infinitesimal in (\ref{bos}) but
often associated with a cutoff of the order of a lattice constant]. 
Renormalization group then allows to derive the gap for 
arbitrary $K_{\sigma}$. The charges remain gapless. 
					   
The Mott insulator is the consequence of an instability 
in the charge channel, caused by Umklapp scattering off the lattice for
commensurate band-fillings. 
The Umklapp Hamiltonian appropriate for a half-filled band is obtained by
simply replacing spin by charge in Eq.~(\ref{h1pbos}), and its
coupling constant often is denoted by $g_{3\perp}$.
Here the spins are gapless while relevant Umklapp scattering opens
a gap $\Delta_{\rho}$ in the charge channel. 
This generic picture applies (with little modification only)
to all even commensurabilities 
($k_F a = [r/s] \pi / 2$, $s$ even). The situation
is different for $s$ odd, where the Umklapp operator necessarily couples
charges and spins \cite{myrev}, and we exclude these cases from our study.
The Mott insulator is dominated
by $4k_F$-CDW and/or SDW correlations. 

While the Luther-Emery solution is essentially \cite{ref} exact, 
it is useless for
computing correlation functions since there is no practical relation
between the physical fermions and the spinless pseudofermions.
Still, we have some qualitative information on the correlation functions. 
Several
methods \cite{lee} support the idea that, in the gapped phase, 
correlations of the 
$\Phi_{\sigma}$-field tend towards a non-zero constant as $| x |$ or $|t|
\rightarrow \infty$ while those involving exponentials of its dual field 
$\Theta_{\sigma}(x)$ decay exponentially in space (or 
oscillate in time). The spin gap quenches
low-energy spin fluctuations, therefore SDW and TS correlations should be
exponentially suppressed. With a constant asymptotic value of 
$\Phi_{\sigma}$, CDW and SS are enhanced with respect to a Luttinger
liquid, and now dominate over SDW and TS. The opening of a
spin gap is a necessary condition for the emergence of dominant SS or CDW
correlations in a 1D metal. As a corollary, a Luther-Emery phase
must exist in the normal state of CDW systems (or superconductors) between 
a Luttinger liquid and the 3D ordered low-temperatures phases. One therefore
should be careful in interpreting the properties of the metallic ``normal
state'' of a CDW system
(or of a 1D superconductor) in terms of a Luttinger liquid.

For the one- and two-particle spectral functions, there 
is a general belief that
the opening of a gap affects the system for frequencies smaller than
this gap while the behavior of the ungapped system is essentially recovered
at larger frequency scales. 
The exponential decay (resp. oscillations) of correlation functions involving
operators $\exp [i(\ldots) \Theta_{\sigma}]$ would cut off (shift) 
the divergences as functions of $q$ ($\omega$) they had possessed
in the Luttinger model. Possibly important power-law prefactors 
to exponentials have not been discussed.
There has been almost no calculation or systematic 
construction of such functions -- in particular dynamical ones \cite{mori} --
and, to my knowledge, no critical check of these hypotheses by numerical
simulation prior to this work \cite{brief1,ucla}.

A wide variety of models fall into the Luther-Emery universality
class and my results should be applicable there in a low-energy
sector: Luttinger liquids coupled to phonons and related models so long as
they are incommensurate, have wide regions of parameter
space with gapped spin fluctuations and gapless charges \cite{zkl}; 
the negative-$U$ Hubbard 
model at any band-filling has a spin gap \cite{negu}, and the positive-$U$ 
Hubbard model at half-filling has a charge gap 
\cite{liebwu,preuss}; with longer-range interactions, charge gaps can
occur at different rational band-fillings, too.
The $t-J$-model has a spin gap at low density \cite{tj}. Spin gaps occur 
frequently in models of two Luttinger or Hubbard chains coupled by
single-particle tunneling
\cite{twoch,tsune}. Also when a $2k_F$-CDW is established in many coupled 
Luttinger chains as a consequence of interchain Coulomb interaction,
the system passes through a region of attractive backscattering which
opens a spin gap \cite{iccoul}.

\section{Construction of the spectral function}
\label{construction}
I now present a systematic construction of the single-particle 
spectral function, Eqs.\ (\ref{specdef}) and (\ref{grefct}), for the
\em spin-gapped \rm Luther-Emery model. 
The Green's function
exhibits the full complexity of the problem, involving all four phase
fields $\Phi_{\nu}$, $\Theta_{\nu}$, while many others are easier
\cite{myrev}. They will be discussed in Section \ref{other}. Here, we
limit ourselves to the diagonal terms of the Green's function, both in the
branch index $r$ and in the spin index $s$, and further assume spin-rotation
invariance, so that $s$ is dropped alltogether. This assumption, which I
will make throughout the paper unless exceptions for the sake of an
argument are stated explicitly, further implies $K_{\sigma} =1$.
With the nonvanishing 
expectation values of operators $\exp[ i (\ldots) \Phi_{\sigma}]$ generated
by the gap opening, finite off-diagonal terms are possible, in principle, 
both here and in multi-particle correlation functions. They
can be calculated in complete analogy to the terms discussed here, and
we ignore them in the following.

Using bosonization (\ref{bos}), the retarded Green's function
for right-moving fermions ($r=+$) can be represented as a product 
\begin{equation}
\label{ggnu}
G(xt) = -i \Theta(t) e^{ik_Fx}  
\left[ g_{\rho}(xt) g_{\sigma}(xt) + (x \rightarrow -x,
t \rightarrow -t) \right] \;\;,
\end{equation}
of charge and spin correlation functions
\begin{equation}
\label{gnu}
g_{\nu}(xt) =  \left\langle \Psi_{+\nu}(xt) \Psi^{\dagger}_{+\nu}(00)
\right\rangle \;\;\;.
\end{equation}
The product structure is
a consequence of the charge-spin separation of the Hamiltonian (\ref{hlutt}).
The spectral function (\ref{specdef}) then is a convolution 
\begin{eqnarray}
\label{specconv}
\rho(q,\omega) & = &  \frac{1}{(2 \pi)^{2}} 
\int_{-\infty}^{\infty} \! d {q'} \:
d {\omega}' \: \left[ g_{\rho}({q}',{\omega}') g_{\sigma}(q-{q}', \omega
-\omega') + (q \rightarrow - q \;, \; \omega \rightarrow - \omega ) \right]
\;\; .
\end{eqnarray}

The charge part is easy and can be calculated in the Luttinger model
(I only display the leading $\omega$- and $q$-dependence)
\begin{eqnarray}
\label{chargecoup}
g_{\rho}(q,\omega) & \sim &  \Theta(\omega
- v_{\rho} q) \Theta(\omega + v_{\rho} q)
(\omega - v_{\rho}q)^{\gamma_{\rho}-1} 
(\omega + v_{\rho} q)^{\gamma_{\rho} -1/2}   \\
{\rm with}  & ~ & \gamma_{\rho} = ( K_{\rho} + K_{\rho}^{-1} - 2) / 8
\;\;\; {\rm for} \;\;\; K_{\rho} \neq 1 \;\;, \nonumber \\
\label{chargefree}
& \sim & \frac{\Theta(\omega + v_{\rho} q)}{\sqrt{\omega + v_{\rho} q}}
\delta(\omega - v_{\rho} q) \hspace{0.5cm} (K_{\rho} = 1) \; . 
\end{eqnarray}

Using a similar expression for the spins, one can reproduce in detail the
spectral functions of the Luttinger model calculated elsewhere directly
\cite{specll,sanseb,sm}. Notice that the divergences are stronger than for 
a spinless Luttinger model ensuring that singularities remain  after
performing the convolution integrals. For both 
$K_{\nu} \neq 1$, the coalescence
of three of the four singularities of $g_{\rho}(q,\omega)$ and
$g_{\sigma}(q,\omega)$ is needed to generate a singularity
in the spectral function of the Luttinger model; if one of them, e.g.
$K_{\sigma}$, is unity, the coalescence of two singularities is sufficient.

The determination of the spin correlation function is more involved because
it has no simple representation in terms of the Luther-Emery pseudofermions,
\em excluding any exact calculation. \rm I now show that the leading behavior 
of this function can,
however, be \em uniquely constructed from symmetries, equivalences, and known 
limits \rm if the Ansatz is made that $g_{\sigma}(xt)$ is a product of
power laws and exponentials in $x$ and $t$. There is a variety of arguments
requiring this form, and we will give them in the following,
together with the construction procedure.

The important steps are: (i) Representing the Hamiltonian in terms of 
right- and
left-moving fermions requires $g_{\sigma}$ to be a function of 
$x \pm v_{\sigma} t$
only. In general, $g_{\sigma}$ will contain both power laws ($f_{\pm}$)
and exponentials ($f_{\rm exp}$)
of these variables 
\begin{equation}
\label{gsigma}
g_{\sigma}(xt) \sim f_+(x-v_{\sigma}t) f_-(x+v_{\sigma}t) 
f_{\rm exp} (x^2 - v_{\sigma}^2 t^2) \;\; .
\end{equation} 
Interactions other than $g_4$ can only mix left- and right-moving excitations,
producing products of $x \pm v_{\sigma} t$, or functions thereof, but cannot
introduce new dependences on $x$ and/or $t$. 
This is consistent both with the boson solution
of the massless Luttinger phase and with the Luther-Emery solution of the
gapped phase. (The Lorentz invariance of the Luther-Emery model requires all
correlation functions of Luther-Emery \em pseudofermions \rm to depend on
$x^2 - v_{\sigma}^2 t^2$ only -- and by implication all those of the 
physical fermions whose operators can be represented in terms on Luther-Emery
fermions alone.)
The exponential part $f_{\rm exp}$ necessarily is a function
of $x^2 - v_{\sigma}^2 t^2$ only.
All dependences on $x$ and $t$ other than through functions of
$x^2 - v_{\sigma}^2 t^2$ must therefore be present also in the Luttinger model
($g_{1\perp} = 0$), and necessarily are of power-law form.

(ii) The limit of a vanishing gap $\Delta_{\sigma} \rightarrow 0$
can also be used to constrain the
function $g_{\sigma}(xt)$, but is rather subtle. 
To make the argument clear, we momentarily relax the assumption of spin
rotation invariance so that the spin channel of the model is described by
$g_{1\perp}$ and general $K_{\sigma}$. (Alternatively, we can look at a
Mott problem with Umklapp scattering $g_{3\perp}$ and $K_{\rho} \neq 1$
is more natural.) In the limit $\Delta_{\sigma} \rightarrow 0$, the function
$f_{\rm exp}(x^2 - v_{\sigma}^2 t^2) \rightarrow 1$ here, because 
the exponential 
dependences are introduced by the finite gap. Straightforwardly, one would
now identify the product $f_+(x-v_{\sigma}t) f_-(x+v_{\sigma}t) =
\langle \Psi_{\sigma} (xt) \Psi_{\sigma}^{\dagger}(00) \rangle_{g_{1\perp}=0}$
with the spin part of the spectral function of the remaining Luttinger
model, i.e. Eq.~(\ref{ffpm}) below with \em anomalous \rm exponents
$\delta_- = (K_{\sigma} + K_{\sigma}^{-1} -2)/8$ and $\delta_+ = \delta_- + 
1/2$. This physically
appealing procedure was used in an earlier paper \cite{brief1}, and
possibly could describe the physics of a small-gap Luther-Emery model.

Taking the limit $\Delta_{\sigma} \rightarrow 0$ to constrain eventual
power-laws in $g_{\sigma}(xt)$ involves different physics, however,
and the above argument must be modified. 
For vanishing gap, $g_{\sigma}$ must reduce to the correlation function
of the \em free \rm Luttinger model ($K_{\sigma} = 1$), 
no matter what value of $K_{\sigma}$ \em would \rm
describe the hypothetical Luttinger model obtained from the Luther-Emery model
(\ref{hlutt}) for $g_{1\perp} = 0$, i.e. independently of any assumption
on spin-rotation invariance. 
Physically, this is so because the anomalous operator
dimensions $K_{\sigma} \neq 1$ of the Luttinger model are a consequence
of singular low-energy virtual particle-hole excitations. When there
is a gap at the Fermi surface, these processes are quenched, and one is
left with the exponent $K_{\sigma} = 1$ of the free model \cite{early}. 
Notice that this argument implies that we consider a rather large gap.

Accidentally, the spectral functions given earlier \cite{brief1}
remain correct. This, however, is due to the limitation to spin-rotation 
invariant interactions there. They impose $K_{\sigma}=1$ for the power-law 
functions $f_{\pm}(x \mp v_{\sigma} t)$ in any case. 

With $f_{\rm exp}(x^2 - v_{\sigma}^2 t^2; \Delta_{\sigma} =0)
\sim 1$ one can determine all possible power-laws $f_{\pm}$ up to 
corrections varying more slowly than a power law, to be
\begin{equation}
\label{ffpm}
f_{\pm}(x \mp v_{\sigma} t) = \left[ \alpha -i (x \mp v_{\sigma} t)
\right]^{-\delta_{\pm}}
\end{equation}
with exponents
\begin{equation}
\delta_+  =  1/2  \;\;, \hspace{0.5cm}
\delta_-  =  0 \;\;.
\end{equation}
These are the exponents
of a free Luttinger correlation function for the spin part of a right-moving 
fermion. I remphasize that they arise because of the quenching of low-energy
particle-hole excitations by the spin gap and hold independent of any
assumption on spin-rotation invariance. (As we will see below, 
the corresponding result for the charge channel implies that there cannot
be any anomalous dimensions in a 1D Mott insulator with spin-rotation
invariance respected).

(iii) From the equivalence of the 
Luther-Emery model to a classical 2D Coulomb gas \cite{chui}
(using the Matsubara
formalism of imaginary times $\tau = it$, putting $y = v_{\sigma} \tau$)
and Debye screening of the charges above the Kosterlitz-Thouless
temperature, one deduces an exponential factor 
\begin{equation}
\label{fexp}
f_{\rm exp} (x \pm v_{\sigma} t) \sim
\exp (- c \Delta_{\sigma} \sqrt{x^2 / v_{\sigma}^2 - t^2}) 
\end{equation}
with an undetermined constant $c$, in 
$f_{\rm exp}$. This equivalence quite generally excludes any decay faster
than (\ref{fexp}). 
 
In this picture, the perturbation Hamiltonian (\ref{h1pbos}) generates
a Coulomb gas of charges $q_e = \pm 1$, and the $\Phi_{\sigma}$-fields
of the Green's function appear as two test charges $q_e' = \pm 1/2$ whose
(bare logarithmic) interaction is modified by screening from the Coulomb gas. 
The gapped Luther-Emery phase corresponds to the high-temperature plasma
phase of unbound charges in the Coulomb gas, 
and the screening can then be treated in the
Debye-H\"{u}ckel approximation \cite{dh}. Here, the effective potential
between the charges is $V(r) \sim \exp(-\kappa_D |\vec{r}|) / \sqrt{\kappa_D 
| \vec{r}|}$ with the Debye wavevector $\kappa_D = 2 \Delta_{\sigma}
/ v_{\sigma}$ \cite{chui}. 
The $\Theta_{\sigma}$-fields can then be viewed as magnetic monopoles
with strengths $q_m = \pm 1/2$. Their interaction is again logarithmic,
and they couple to the electric charges with $V_{em}(\vec{r}) \sim - \arctan
(y/x)$ \cite{kada}. Clearly, the high-temperature plasma of electric charges
$e_e = \pm 1$  modifies the effective monopole-monopole
interaction which becomes
\begin{equation}
\label{vmm}
V_{\rm m-m}(\vec{q}) = - \frac{2 \pi}{q^2} + \frac{2 \pi}{q^2} 
\left( \frac{q_y}{q_x} \right)^2 \frac{1}{q^2 + \kappa_D^2} \;\;\;,
\end{equation}
where I have used the Debye-H\"{u}ckel polarization propagator
\begin{equation}
\Pi(\vec{q}) = \frac{q^2}{2 \pi} \frac{\kappa_D^2}{q^2 + \kappa_D^2} \;\;\;.
\end{equation}
Fourier-transforming back to real space, one obtains
\begin{equation}
\label{vrr}
V_{\rm m-m}(\vec{r}) \sim \ln | \vec{r} | + c' | \vec{r} | \kappa_D
\end{equation}
with an open constant $c' \propto c$. One observes an antiscreening
effect here: in the presence of the electric charges, the magnetic monopoles
are confined more strongly than without charges!
Going back to real times, (\ref{vrr})
produces the exponential dependence in (\ref{fexp}) and, most
importantly, gives additional (in fact, for those
multi-particle correlation functions which only depend on $x^2 - v_{\sigma}^2 
t^2$ the only firm) justification for 
the presence of power-law prefactors in 
addition to exponential terms in (\ref{gsigma}).

(iv) The open constant $c$ in (\ref{fexp}) can be determined
from a spectral representation of $f_{\rm exp}$, and our interpretation
of the bosonization formula (\ref{bos}). Fourier transforming $f_{\rm exp}
(x,t)$, one obtains 
\begin{equation}
f_{\rm exp}(q,\omega) = 2 \pi v_{\sigma} c \Delta_{\sigma}
\frac{\Theta(\omega^2 - v_{\sigma}^2 q^2 - c^2 \Delta_{\sigma}^2) }{
(\omega^2 - v_{\sigma}^2 q^2 - c^2 \Delta_{\sigma}^2)^{3/2}}
\end{equation}
which has a gap of magnitude $c \Delta_{\sigma}$ in its spectrum. 
This gap must correspond to the excitation of $|n|$ spinons 
where $n$ is the prefactor of $i \Theta_{\sigma}(x) /
\sqrt{2}$ in the operator whose correlation function we wish to calculate.
This constrains the prefactor in the exponential to $c = |n|$ quite 
generally. For the single-particle Green's function $n = 1$, and we
obtain $c=1$ here.

(v) The present construction of $g_{\sigma}(xt)$ is not an exact calculation.
It is therefore important to look for exactly known cases which can be
used as tests, to confirm the validity of this construction. Gul\'{a}csi has
calculated explicitly the $t=0$-Green's function of a 1D Mott insulator
\cite{miklos}: He finds $G(x) \sim \exp(- \Delta_{\rho} |x|) / |x|$ which 
is in complete agreement with the present theory when the 
$1/\sqrt{|x|}$-contribution from the ungapped channel is multiplied to
Eq.\ (\ref{spreal}) below. That there may be a power-law prefactor in the
charge part of the spectral function 
has also been realized but well hidden in publications,
by others \cite{stamil}.

In Section \ref{other}, I will discuss further tests of these
rules based on two-particle correlation functions.

From the rules (i) -- (v), I  find 
\begin{equation}
\label{spreal}
g_{\sigma}(x,t) \sim 
\exp \left( - \Delta_{\sigma} \sqrt{x^2 - v_{\sigma}^2 t^2} 
/ v_{\sigma} \right) / \sqrt{\alpha + i (v_{\sigma}t - x)} \; .
\end{equation}

Fourier transformation then gives
\begin{equation}
\label{spksp}
g_{\sigma}(q,\omega) \sim \left( 1 +
\frac{v_{\sigma} q}{\sqrt{v_{\sigma}^2 q^2 + \Delta_{\sigma}^2}} \right) \:
\frac{\Theta(\omega + v_{\sigma} q)}{\sqrt{\omega + v_{\sigma} q}} \:
\delta(\omega - \sqrt{v_{\sigma}^2 q^2 + \Delta_{\sigma}^2}) \;\;.
\end{equation}

The comparison of (\ref{spksp}) with (\ref{chargefree}) (after $\rho
\rightarrow \sigma$ there) is interesting. The $\delta$-function translates
the absence of anomalous dimensions in the gapped channel, a consequence
of rule (ii), rather than spin-rotation invariance as in the 
$\sigma$-version of (\ref{chargefree}). The change in dispersion due to
the spin gap enters through this $\delta$-function. The frequency-dependent
prefactor is the same as in the gapless system. However, due to the
different argument in the $\delta$-function, it no longer becomes singular
in the limit $q,\omega \rightarrow 0$ but has an upper limit of
$\Delta_{\sigma}^{-1/2}$ now. A similar effect occurs in the Green's function
of 1D quantum antiferromagnets, where the opening of the spin gap cuts off a
singularity of the prefactor of the delta function \cite{brenig}.
The factor in parentheses is a coherence factor translating the enhanced
spin-pairing tendency at the origin of the spin gap, and one readliy 
recognizes the same structure as for the coherence factors $u_q , \;
v_q$ familiar from the theory of superconductivity.

\section{Spectral function for the spin-gapped Luther-Emery model}
\label{specfle}
We now must convolute $g_{\sigma}(q,\omega)$, Eq.~(\ref{spksp}), 
with the charge part, Eqs.~(\ref{chargecoup}) or (\ref{chargefree}). 
The results depend on the relative magnitudes of the charge and spin
velocities. We therefore treat separately the cases of (A) repulsive
interactions (in the sense that the effective forward scattering matrix
element $g_{2 \rho} - g_{1\|}/2 >0)$, i.e. $K_{\rho} < 1$, and 
$v_{\rho} > v_{\sigma}$, where Peierls-type $2k_F$-CDW fluctuations
dominate, and (B) attractive forward scattering, i.e. $K_{\rho} > 1$
and $v_{\rho} < v_{\sigma}$, when singlet superconducting fluctuations
are most important. (The inequalities on the velocities and $K_{\rho}$ 
usually go with each other as listed
when standard lattice models are treated. Of course, when one takes all
$g_{i \nu}$ as free parameters, other combinations are possible.
Relevant for the subsequent classification then are the velocities.) 

What could we expect from our knowledge of the Luttinger liquid \cite{specll}?
There the singularities at $\omega = v_{\rho (\sigma)} q$ 
arise from processes where the charge (spin) contributes all of the electron's
momentum $q$ and the spin (charge) none. The same argument applied to the
Luther-Emery model predicts signals at the renormalized spin dispersion
$\varepsilon_{\sigma}(q)$, Eq.\ (\ref{esigma}),
and at a shifted charge dispersion 
\begin{equation}
\label{erho}
\varepsilon_{\rho}(q) = v_{\rho} q + \Delta_{\sigma} \;\;.
\end{equation}
Figure 1 shows the location of the signals expected from this argument.
The $\Delta_{\sigma}$-shift in the charge dispersion comes
from the fact that the zero-momentum spin fluctuation can only be excited
at a cost of $\Delta_{\sigma}$. As will be seen below, however, the spectral
functions of the Luther-Emery model never show two singularities with these
dispersions. The intuitive predictions on the spectral function of the 
Luther-Emery model basically transcribe the standard argument that the 
behavior of correlation functions is modified on energy scales below the gap
(correlations are suppressed there) but recovered almost unchanged on higher
energy scales. Our results will show that for dynamical, $q$ \em and \rm
$\omega$-dependent correlations, this argument is not trustworthy.

\subsection{One-dimensional Peierls ``insulators''}
We assume $v_{\rho} > v_{\sigma}$ and $K_{\rho} < 1$,
implying dominant CDW correlations. Calling these systems ``insulators''
is a misnomer, however, because the charges are gapless and the systems
are metallic. More precisely, we think about the Luther-Emery model here
as describing the ``normal'' metallic state above a CDW transition.

The convolution of $g_{\sigma}$ and $g_{\rho}$, Eq.\ (\ref{specconv}),
is rather straightforward now. After executing the $\omega'$-integral,
singularities are obtained from the
coalescence of the two singularities carried by $g_{\rho}(q,\omega)$.
The result of the calculation is shown schematically in Fig.\ 2
for $q < 0$ (unlike previous papers, we present the 
spectral functions as those of
the \em occupied \rm states, i.e. as they would be measured by a 
photoemission experiment). There are indeed features at the special
frequencies shown in Fig.\ 1. On the spin dispersion $\varepsilon_{\sigma}(q)$, 
there is a true singularity 
\begin{equation}
\label{specsig}
\rho[q,\omega \approx -\varepsilon_{\sigma}(q)]
\sim \Theta[- \omega - \varepsilon_{\sigma}(q)]
[-\omega - \varepsilon_{\sigma}(q)]^{\alpha-1/2}
\end{equation}
as in the Luttinger model. Here, $\alpha$ is \em defined \rm as 
$\alpha = (K_{\rho}+K_{\rho}^{-1}-2)/4 = 2 \gamma_{\rho}$ 
since the notion of a $K_{\sigma}$ does not make sense in a spin-gapped
system. 
Folklore would then 
predict another singularity
$| \omega + \varepsilon_{\rho}(q)|^{(\alpha-1)/2}$ (short dashed lines in
Fig.\ 2) 
which is \em not \rm observed here. It is cut off instead to a 
finite maximum of order 
\begin{equation} 
\rho[q,\omega \approx - \varepsilon_{\rho}(q)]
\sim \Delta_{\sigma}^{(\alpha-1)/2} \;\;.
\end{equation}
The reason for cutting of the Luttinger divergence on the charge dispersion
is related to the non-singular prefactor (for $q \rightarrow 0$) in 
$g_{\sigma}(q,\omega)$, cf.\ Eq.\ (\ref{spksp}) and the subsequent discussion,
and the convolution makes this 
effect apparent on the charge dispersion $\varepsilon_{\rho}(q)$. 
\em The spin gap therefore supresses the divergence
associated with the charge dispersion while on the renormalized spin 
dispersion, the spectral response remains singular. \rm  

At positive frequencies, the Luther-Emery model has pronounced shadow bands.
Here, the Luttinger liquid only has very small weight. 
The weight in the Luther-Emery model
is much stronger, and the spectral function has the same overall shape
as at negative frequencies. For $q<0$, the negative frequency part is 
enhanced by a coherence factor 
$1 - v_{\sigma} q / \varepsilon_{\sigma}(q)$
while a factor $1 + v_{\sigma} q / \varepsilon_{\sigma}(q)$ decreases its
shadow. These factors translate the increased coherence due to the 
spin pairing and the finite spin gap, and are a consequence of the 
corresponding coherence factors in Eq.\ (\ref{spksp}). Of course, as
suggested by Fig.\ 1, one can also view the shadow bands as bending back
from the Fermi (or more precisely: the gap)
energy as $k$ is increased beyond $k_F$. This view perhaps
is closer to a real photoemission experiment.

\subsection{One-dimensional superconductors}
We now take $v_{\rho} < v_{\sigma}$, i.e. attractive forward scattering.
This implies $K_{\rho} > 1$, and such a system has dominant singlet
pairing fluctuations. Interestingly, two true singularities occur
here whose location is shown in Fig.\ 3. 
There is one singularity on the renormalized spin dispersion
\begin{eqnarray}
\rho[q, \omega \approx - \varepsilon_{\sigma}(q)] & \sim &
\Theta(q-q_c) \Theta[-\omega - \varepsilon_{\sigma}(q)]
[- \omega - \varepsilon_{\sigma}(q)]^{\alpha-1/2} \nonumber \\
& + &\Theta(q_c - q) |- \omega - \varepsilon_{\sigma}(q)|^{\alpha-1/2} \;\;,
\end{eqnarray}
which is one-sided for $q > q_c$ and two-sided
for $q < q_c$.  
\begin{equation}
q_c = {\rm sign} (q) \frac{v_{\rho}}{v_{\sigma}} \frac{\Delta_{\sigma}}{
\sqrt{v_{\sigma}^2 - v_{\rho}^2}}
\end{equation}
is a critical wave vector which arises in the convolution
procedure from searching the minimum of $\varepsilon_{\sigma}
(q') + v_{\rho}(q - q')$ 
as a function of $q'$. At this wavevector, the dispersion 
\begin{equation}
\tilde{\varepsilon}_{\rho}(q) = \varepsilon_{\sigma}(q_c) + v_{\rho} (q-q_c)
\end{equation} 
is tangential to $\varepsilon_{\sigma}(q)$. For $q < q_c$, a divergence
\begin{equation}
\rho[q, \omega \approx - \tilde{\varepsilon}_{\rho}(q)] \sim 
\Theta(-\omega - \tilde{\varepsilon}_{\rho}(q))
[-\omega - \tilde{\varepsilon}_{\rho}(q)]^{\frac{\alpha - 1}{2}}
\end{equation}
on this shifted charge dispersion splits off the spin divergence.
Again, there are strong shadow bands with the same functional forms as
the main bands, specifically with two singularities, and with intensities
controlled by coherence factors. The dispersions 
of the signals are displayed in Figure 3, 
and the shape of the spectral function is sketched in Figure 4.

Notice that, quite generally, that the behavior of $\rho(q,\omega \approx \pm
\Delta_{\sigma})$ is determined
by that of the spin part close to $\Delta_{\sigma}$ and that of the charge part
at $\omega \approx 0$. Unlike earlier conjectures \cite{lee}, 
it is therefore \em not \rm necessary to know details
of the charge dynamics on a scale $\omega \approx \Delta_{\sigma}$ where
the Luttinger description may have acquired significant corrrections. 

The $k$-integrated density of states then is $N(\omega) \sim
\Theta(\omega - |\Delta_{\sigma}|)  (\omega - |\Delta_{\sigma}|)^{\alpha}$,
independent of the magnitudes of the velocities. 
There is no weight below the gap, and the
typical gap singularity in the density of states of the spin 
fluctuations is wiped out by convoluting with the gapless charges. 

It is quite clear now that certain properties of 1D fermions -- 
the dynamical ones involving (1+1)D Fourier transforms
-- are affected by the gap opening on \em all \rm energy
scales, contrary to common expectation,
while those depending on one variable alone are modified only on
scales below the gap energy. Despite the opening of a gap in the 
spin channel, singular spectral response
remains possible in $q$- and $\omega$-dependent correlation functions.

\section{Spectral function of one-dimensional Mott insulators}
\label{specfmott}
The spectral function of a 1D Mott insulator can be computed as a special
case of the generic solution presented above. One simply has to change
$\sigma \leftrightarrow \rho$ \em everywhere \rm 
and put $K_{\sigma} = 1$ in the
gapless spin channel for spin-rotation invariance (which we assume to hold, 
again). Importantly, the
exchange of $\rho$ and $\sigma$ also applies to the inequalities on
the velocities $v_{\nu}$, where again two cases must be distinguished. 

Both factors $g_{\nu}$ in the convolution now involve $\delta$-functions.
In the case of repulsive forward scattering $v_{\rho} > v_{\sigma}$,
one now finds a spectral function with two singularities, similar to the
case of a 1D superconductor. Since $K_{\sigma} = 1$, the anomalous
single-particle exponent $\alpha =0$, i.e. one obtains two inverse 
square-root singularities. In the main band ($\omega < 0$ for $q<0$),
the spectral function becomes
\begin{equation}
\rho(q,\omega) \sim \frac{\Theta(q-q_c) \Theta[-\omega - \varepsilon_{\rho}(q)]
}{\sqrt{-\omega - \varepsilon_{\rho}(q) }}
 +  \frac{\Theta(q_c - q) 
\Theta[-\omega - \tilde{\varepsilon}_{\sigma}(q)]}{
\sqrt{[-\omega - \tilde{\varepsilon}_{\sigma}(q)] 
| -\omega - \varepsilon_{\rho}(q) |}}
\end{equation}
with $\tilde{\varepsilon}_{\sigma}(q) = \varepsilon_{\rho}(q_c) + v_{\sigma}
(q-q_c)$. 
An important difference to the case of a superconductor occurs in the
shadow band: since the spectral function of the gapless spin channel
has no shadow band of its own, the singularity on 
$\tilde{\varepsilon}_{\sigma}(q)$ in the shadow band is missing. The shadow band
therefore has a single
singularity on the charge dispersion $\varepsilon_{\rho}(q)$
with a weight depressed by a coherence factor with respect
to the weight of the main band signals. The effect is completely 
analogous to the appearence of a single nonanalyticity in the (very weak)
shadow bands of a Luttinger liquid with spin-rotation invariant interactions
\cite{specll,sanseb,sm}. The shape of this spectral function is sketched
in Figure 5. The location of the singularities follows Figure 3 with 
the replacement $\rho \leftrightarrow \sigma$ except for the shadow bands
where the straight lines should be ignored. 

The case $v_{\sigma} > v_{\rho}$ again is different. Compared to the 
case of the 1D Peierls ``insulator'', the anomalous dimension $\alpha$ on the
charge dispersion drops out due to spin-rotation invariance, giving an
inverse square-root singularity on $\varepsilon_{\rho}(q)$. Also the
finite maximum on the shifted spin-dispersion $\varepsilon_{\sigma}(q)$
does not occur. This is because the $\delta$-function has zero weight
in the energy domain where the square-root prefactor in Eq.\ (\ref{spksp})
takes its maximum. The shadow band, of course, has a single 
inverse-square-root singularity with the usual coherence factors.
Thus, the spectral function for this case becomes
\begin{equation}
\rho(q,\omega) \sim \frac{\Theta[|\omega| - \varepsilon_{\rho}(q)] }{
\sqrt{|\omega| - \varepsilon_{\rho}(q)}} \;\;,
\end{equation}
up to coherence factors, and the density of states
\begin{equation}
N(\omega) \sim \Theta(|\omega| - \Delta_{\rho}) \times 
{\rm regular \; function} \;. 
\end{equation}

The spectral properties of a doped Mott insulator, of course, depend on
the detailed scenario emerging from a more complete theory. Work on the
Hubbard model shows, however, that the upper Hubbard band qualitatively
survives a finite dopant concentration \cite{preuss,miklos}. 
Continuity then suggests that as the insulating state is left by
varying the band-filling, spectral weight is
gradually taken out of both the main and shadow bands of a spectral
function such as those discussed before, and transferred into 
the charge and spin divergences of a Luttinger
liquid signal. Although the spins are left unaffected in the transition
and only a charge gap opens,
both the charge and the spin signals are predicted to be
shifted and strongly modified by doping. This is a direct consequence of the 
convolution property (\ref{specconv}) of the single-particle spectral 
function. When superposing (to a first 
approximation) the two signals, care must be taken, in addition, to 
account for the dependence of the chemical potential on doping level. 

\section{Generalization to other correlation functions}
\label{other}
We now discuss the construction of other correlation functions
for the Luther-Emery model. Clearly, due to charge-spin separation,
they can again be written as convolutions of charge and spin correlation
functions. 
Consider a general local operator 
\begin{equation}
O_{\nu}^{(m,n)}(x) = \Psi_{r\nu}^m(x) \Psi_{r\nu}^n(x) \;\;\;,
\end{equation}
where $\Psi_{r\nu}(x)$ had been introduced in Eq.\ (\ref{psinu}), and
a positive (negative) exponent is understood as a creation (annihilation)
operator. Bosonizing $O_{\nu}$, the $\Phi_{\nu}$-field acquires
a prefactor $(m-n)$, and $\Theta_{\nu}$ is multiplied by $(m+n)$ with
respect to the single-particle operator $\Psi_{r\nu}$. If gapless
channel is assumed to be the charge $\nu= \rho$ (as we have done throughout
this paper except in the preceding section), 
the correlation function of $O_{\rho}$ behaves as
\begin{eqnarray}
R^{(m,n)}_{\rho}(xt) & = &
\left\langle O_{\rho}^{(m,n)}(xt) \left[O_{\rho}^{(m,n)}(00) \right]^{\dagger}
\right\rangle_{\rm Luttinger} \nonumber \\
& \sim & \left( \alpha + i v_{\rho}
t - ix \right)^{-m^2/2}
\left( \alpha + i v_{\rho} t + ix \right)^{-n^2/2}  \nonumber \\
& \times & \left( \frac{\alpha^2}{
(\alpha + i v_{\rho}t )^2 + x^2} \right)^{\frac{(m-n)^2}{8}(K_{\rho}-1)
+ \frac{(m+n)^2}{8}(K_{\rho}^{-1}-1)}
\;\;.
\end{eqnarray}
Its Fourier transform is 
\begin{eqnarray}
\label{mnrho}
R^{(m,n)}_{\rho}(q,\omega) & \sim &  \Theta(\omega
- v_{\rho} q) \Theta(\omega + v_{\rho} q)
(\omega - v_{\rho}q)^{\gamma_-^{(m,n)}-1} 
(\omega + v_{\rho} q)^{\gamma_+^{(m,n)}-1} \;\;,  \\
\gamma_+^{(m,n)} & = & \frac{m^2}{2} + \frac{(m-n)^2}{8} (K_{\rho} -1)
+ \frac{(m+n)^2}{8} \left( \frac{1}{K_{\nu}} - 1 \right) \;\;, \nonumber \\
\gamma_-^{(m,n)} & = & \frac{n^2}{2} + \frac{(m-n)^2}{8} (K_{\rho} -1)
+ \frac{(m+n)^2}{8} \left( \frac{1}{K_{\nu}} - 1 \right) \;\;. \nonumber 
\end{eqnarray}

We now turn to such an operator for spin, $O_{ \sigma}$, in the presence
of a spin gap.
When the spin gap opens due to the Hamiltonian (\ref{h1pbos}), 
the $\Phi_{\sigma}$-field develops long-range order. Its dual field, 
$\Theta_{\sigma}$, then is disordered, and its correlations will contain
exponential terms similar to $f_{\rm exp}$, Eq.\ (\ref{fexp}). 
We now have to distinguish two cases.
(i) If we have $m = -n$, the operator $O^{(m,-m)}_{\sigma}$ can 
be represented in terms of the $\Phi_{\sigma}$-field alone. 
Since this is the ordering field, we simply can put it to a constant value, 
implying $R_{\sigma}^{(m,-m)}(xt) \sim 1$, 
and the space-time dependence of the total correlation function is then 
determined by the charge part $R_{\rho}^{(s,t)}(q,\omega)$ (which may
carry different powers $s,t$
of $\Psi_{r,\rho}$, depending on the spin directions) alone, and
given by Eq.\ (\ref{mnrho}). One can, in principle, go one step further and
account for the long-wavelength fluctuations out of the ground state-value of
$\Phi_{\sigma}$. A convenient method for this again is the mapping onto
a classical 2D Coulomb gas. Since the $\Phi_{\sigma}$-field of the correlation
functions introduces electric test charges, we know that in the massive
Luther-Emery phase their interaction is exponentially screened
(cf.\ Section \ref{construction}). We then 
find the fluctuation contribution 
\begin{equation} 
\langle [ \Phi_{\sigma}(xt) - \Phi_{\sigma}(00) ]^2 \rangle \sim
\frac{\exp \left(- 2 \Delta_{\sigma}\sqrt{x^2/v_{\sigma}^2 - t^2}\right) }{
(x^2/v_{\sigma}^2 - t^2)^{1/4}} \;\;\;.
\end{equation}
I will discuss an interesting application in a moment.

However, if (ii) $m \neq -n$,
the spin correlations contain the disorder field $\Theta_{\sigma}$
dual to the $\Phi_{\sigma}$, and the gap opening will lead to exponential
factors as in Eq.\ (\ref{fexp}). This is the case
for the Green's function, cf. Eq.\ (\ref{gnu}). 
We apply the same rules (i) -- (v) as in Section \ref{construction}. 
Specifically, the prefactor of the gap in the exponential is $c = |m+n|$,
by comparing the energy for the insertion of $|m+n|$ $\sigma$-particles
into the system with the gap obtained in the spectral representation of
the exponential. The power-law prefactor is that of the free Luttinger
model because there cannot be any anomalous dimensions in a gapped fermion
system. In $(xt)$-space, the correlation function then is
\begin{eqnarray}
R^{(m,n)}_{\sigma}(xt) & = &
\left\langle O_{\sigma}^{(m,n)}(xt) \left[O_{\sigma}^{(m,n)}(00) 
\right]^{\dagger} 
\right\rangle \nonumber  \\
\label{rmnsig}
& \sim & \left( \alpha + i v_{\sigma}
t - ix \right)^{-m^2/2}
\left( \alpha + i v_{\sigma} t + ix \right)^{-n^2/2}  
\exp \left( - |m+n| \Delta_{\sigma} \sqrt{x^2 / v_{\sigma}^2 - t^2}\right)
\;\;.
\end{eqnarray}
This expression can be Fourier transformed and convoluted with
an appropriate charge part.

What the present construction cannot do, however, is to give information
on the magnitude, or a possible vanishing, of the prefactor of the correlation
function. One example is the $2k_F$-CDW correlation function in the half-filled
replusive Hubbard model, where a naive use of the construction above would 
predict (in real space at $t=0$)
a dependence $\sim x^{-1}$ which, on physical grounds, is not expected
to be important in that model \cite{luthem}. Qualitative information can be
obtained in that situation from renormalization group studies, where one can 
monitor how the amplitude of a correlation function changes as one moves
\em away \rm from a Luttinger liquid fixed point \cite{jpl}. 
A complete solution of this
problem presumably would require an exact boson representation of the 
\em physical \rm fermions in a Luther-Emery model, including fermion raising
operators.

To conclude this Section, I discuss two more test cases for my
construction procedure.  Consider the
transverse $2k_F$-spin-correlation functions \cite{myrev,lee}
\begin{equation}
R_{SDW \perp}(xt) = \langle O_{SDW\perp}(xt) O^{\dagger}_{SDW \perp}(00)
\rangle =
\langle \Psi_{-\downarrow}^{\dagger}(xt) \Psi_{+\uparrow}(xt)
\Psi_{+,\uparrow}^{\dagger}(00) \Psi_{-\downarrow}(00) \rangle 
\end{equation}
in the Luther-Emery spin-gap regime.
The spin density wave operator can also be represented as
\begin{equation}
O_{SDW \perp}(xt) = \frac{e^{2ik_Fx}}{2 \pi \alpha} \exp \left\{
-i \sqrt{2} \left[ \Phi_{\rho}(x) + \Theta_{\sigma}(x) \right] \right\}
= e^{2ik_Fx} O_{\rho}^{(-1,1)}(xt) O_{\sigma}^{(-1,-1)}(xt) \;\;.
\end{equation}
We now limit ourselves to the spin component of the correlation function
and obtain, using Eq.\ (\ref{rmnsig}),
\begin{equation}
R^{(-1,-1)}_{\sigma}(xt) \sim \frac{ \exp \left( - 2 \Delta_{\sigma} 
\sqrt{x^2 / v_{\sigma}^2 - t^2}\right) }{\sqrt{x^2 - v_{\sigma}^2 t^2 }}
\;\;.
\end{equation}
Fourier transformation gives
\begin{equation}
\label{r11}
R^{(-1,-1)}_{\sigma}(q,\omega) \sim \frac{\Theta(
\omega^2 - v_{\sigma}^2 q^2 - 4 \Delta_{\sigma}^2)}{\sqrt{
\omega^2 - v_{\sigma}^2 q^2 - 4 \Delta_{\sigma}^2 }} \;\;.
\end{equation}
One the other hand, \em on the Luther-Emery line \rm $K_{\sigma} = 1/2$, one
can refermionize the operator
\begin{equation}
O_{\sigma}^{(-1,-1)}(x) = \sqrt{2 \pi \alpha} \Psi_-^{\dagger}(x)
\Psi_+^{\dagger}(x)
\end{equation}
in terms of spinless fermions $\Psi_r(x)$, by inverting the spinless
variant \cite{myrev} of the bosonization formula (\ref{bos}).
The limitation of this procedure to the Luther-Emery line is inessential
because different coupling constants will only affect the magnitude of
the spin gap but not the form of the excitation spectrum, so long as
$\Delta_{\sigma}>0$. Now, one can calculate $R_{\sigma}^{(-1,-1)}(q,
\omega)$ as the pairing correlation function of spinless fermions \em
in a fermion representation. \rm Such a calculation has been outlined
by Lee \cite{lee}, and the result derived from his expressions agrees
with Eq.\ (\ref{r11}) both concerning the regions of nonvanishing spectral
weight, and the critical exponents of the singularities. 
Incidentally, my own expressions are more complicated than Lee's
by additional terms and additional occupation functions $n(k)$ and
$1 - n(k)$. They conspire with the coherence factors $[1 \pm v_{\sigma} q
/ \varepsilon_{\sigma}(q)]$ to produce a prefactor $v_{\sigma}^2 q^2 
/ (v_{\sigma}^2 q^2 + 4 \Delta_{\sigma}^2)$ to the leading inverse-square-root
singularity which vanishes as $q \rightarrow 0$. At $q=0$, a subleading
term $\propto \Theta(|\omega| - 2 \Delta_{\sigma})$ times a regular function
remains. Apart these subtle prefactors, the exact fermionic calculation
reproduces the result of the construction procedure advocated here for
the correlation functions of the Luther-Emery model.

A final test is provided by the charge correlations of a 1D Mott insulator.
In general, the charge density operator $\hat{n}(x)$ has contributions at 
wavevectors $q \approx 0, \; 2k_F, \; 4k_F$, etc.
\begin{eqnarray}
\label{cdens}
\hat{n}(x) & \sim & - \frac{\sqrt{2}}{\pi} \frac{\partial \Phi_{\rho}(x) 
}{\partial x} 
 +  \frac{1  }{\pi \alpha} 
\exp \left\{-2ik_Fx + \sqrt{2} i \Phi_{\rho}(x)
\right\} \cos[\sqrt{2} \Phi_{\sigma}(x) ] \nonumber \\
& + & \frac{2}{(2 \pi \alpha)^2} 
\exp \left\{ -4ik_Fx +\sqrt{8} i\Phi_{\rho}(x) \right\} \;\;\;. 
\end{eqnarray}
In a half-filled band, $4k_F = 2 \pi / a$, a reciprocal lattice vector
so that the $4k_F$-term effectively does not oscillate when measured on the
lattice sites. When the Mott gap $\Delta_{\rho}$ opens, the field $\Phi_{\rho}$
orders at a finite constant value. The third term in (\ref{cdens}) then 
translates the long-range charge order, the first term measures the 
long-wavelength fluctuations out of this ordered ground state, and the
second term measures $2k_F$ charge fluctuations. Using the arguments at
the beginning of this section (after $\sigma \leftrightarrow \rho$), 
we obtain from the first two terms
a spectral function
\begin{equation}
\label{rrr}
R_{n}(q,\omega) \sim \delta(q) \delta(\omega) + q^2 \frac{\Theta(\omega^2 
- v_{\rho}^2 - 4 \Delta_{\rho}^2)}{\omega^2 
- v_{\rho}^2 - 4 \Delta_{\rho}^2} \;\;\;.
\end{equation}
The zero-frequency $\delta$-function comes from the ``$4k_F$''-part, and the
high-frequency signal from the $\partial \Phi_{\rho} / \partial x$-term.
In principle, one could also calculate the $2k_F$-part. However, experience
with the Hubbard model suggests that prefactors not specified here suppress
the $2k_F$-CDW fluctuations on the lattice sites \cite{myrev}, and we do not
consider them here (similar, and nonvanishing contributions, however appear
in $2k_F$-SDW correlation functions, and in a ``bond order wave'' which is
best described as a $2k_F$-CDW centered midway between two sites). 

The spectral function $R_{n}(q,\omega)$ has been calculated recently by
Mori and Fukuyama \cite{mori}. They do not give an explicit expression
which would allow to check the critical exponents, but the region of nonvanishing
spectral weight, and the overall shape of the high-frequency signal are 
consistent with Eq.\ (\ref{rrr}), whereas the $\delta$-function in 
Eq.\ (\ref{rrr}) seems to be
missing. It is present, however, in a numerical diagonalization of an
extended Hubbard model \cite{peste}, and provides another, though more 
superficial test of our construction.

\section{Relation to other work}
\label{colleagues}
In the preceding sections, we have discussed some tests for the dynamical
correlation functions of the Luther-Emery model constructed here
\cite{lee,miklos}. Independent verification comes from work on many models
which fall into the Luther-Emery universality class.

In particular, numerical studies have attempted to look into the
spectral properties of correlated fermion models. Quantum Monte Carlo
simulations of the 1D Hubbard model at half-filling, a prototypical 
Mott insulator with $v_{\rho} > v_{\sigma}$ provides evidence for 
pronounced shadow bands, much stronger than those of the doped systems
which form Luttinger liquids \cite{preuss}. At present, the resolution
is not good enough to directly visualize the two dispersing 
inverse-square-root singularities found here. However, recent improvements
on doped Hubbard models \cite{zacher}
lend hope that Quantum Monte Carlo will be able, 
in the near future, to confirm the predicitons made here. 

The 1D $t-J$-model
at half-filling also forms a Mott insulator with $v_{\rho} > v_{\sigma}$, 
and exact diagonalization
of lattices up to 22 sites has allowed a calculation of the spectral 
function of this model \cite{kim}. While the location of regions of finite
spectral weight, and of the singularities agrees with the present study,
numerical diagonalization on such small systems does not allow to determine
the critical exponents of the divergences of the 1D Mott insulator. 

Spin gaps also arise in many lattice models. E.g. for two coupled 
Luttinger, Hubbard, or $t-J$-chains, there are wide regions of parameter
space where the spin fluctuations are massive, and the single-particle
spectral function has been calculated occasionally \cite{tsune}. Again,
exact diagonalization finds important shadow bands \cite{tsune} but the
resolution is not good enough to separate the two dispersing divergences
found in Section \ref{specfle} for a superconductor, not to speak of 
the much weaker signal on the shifted charge dispersion 
$\varepsilon_{\rho}(q)$ predicted above for a CDW system.

Evidence for such a weak signal, and for a divergent signal on the gapped
spin dispersion $\varepsilon_{\sigma}(q)$ comes, however, from exact 
diagonalization of a $t-J-J'$-model where a spin gap opens for certain
values of $J'$ \cite{tohy}. These authors observe a very strong
spinon signal, the holon peak is anomalously weak, as predicted here.

A Bethe Ansatz calculation of spectral functions for a 1D Mott insulator
has recently been performed by Sorella and Parola (SP) based on the 1D
supersymmetric $t-J$ model \cite{sorella}, and also 
confirms essential aspects of the 
present work. In their model, $v_{\rho} < v_{\sigma}$ so that we predict
a single inverse-square-root singularity on $\varepsilon_{\rho}(q)$.
Such a singularity is also found from the Bethe Ansatz solution used by 
SP. When a finite magnetization is included, SP find critical exponents
which explicitly depend on the momentum of the hole created. One would expect
from universality and the possibility to transform a positive-$U$ Hubbard model
into one with negative $U$ by a particle-hole transformation on one spin 
species alone, that such spectral functions should also describe spin-gapped
systems with $v_{\rho} > v_{\sigma}$. We do not find
such momentum-dependences in the work presented here. SP's method, however,
requires the calculation of the ground state and low-energy properties
of the spin Hamiltonian at finite total momentum of the spin system. 
These explicitly
depend on the momentum, and produce the momentum-dependent exponents.
In the Luther-Emery model, one calculates a spinon excitation with
some momentum with respect to a zero momentum ground state.
The momentum-dependent correlation
exponents found by SP certainly are beyond scope and
possibilities of the present model. On the other hand, their method does
not allow to look into more subtle features than critical exponents, such 
as the finite maximum which we found in this case.

\section{Applications to experiments}
\label{appl}
Importantly, our results could prove useful in the description of the 
photoemission properties of certain quasi-1D materials. 

There have been angle-resolved photoemission experiments on the 1D
Mott insulator $Sr Cu O_2$ with a gap $2 \Delta_{\rho} \sim 1.8 eV$
\cite{kim}. The lineshapes observed were
anomalously broad and showed unsual dispersion. As a consequence, the
authors proposed a description in terms of a system with charge-spin 
separation, where the broad feature would, in fact, be composed of the
unresolved spin and charge signals. In addition, a strong shadow
band bends back from the gap edge for $k>k_F$. Its
dispersion is consistent with the one of the charge signal
for $k<k_F$. Clearly, these observations are fully consistent with the
theory presented here, which predicts two inverse-square-root singularities
beyond some critical wave vector (cf.\ Fig.\ 5), and a single one below, 
as are the
accompanying diagonalization results on a 1D $t-J$-model \cite{kim}.

More interesting in the present context are a number of unexplained 
ARPES results on organic and inorganic materials which undergo Peierls
transitions at low temperatures. 
Specifically, ARPES experiments on the blue bronze 
$K_{0.3} Mo O_3$ by several groups show
\em two \rm dispersing peaks \cite{gweon}.
Also in the organic conductor $TTF-TCNQ$, anomalous lineshapes are observed
\cite{ttf}.
Of interest here is the $TCNQ$-band which shows $2k_F$-CDW fluctuations
in the metallic state \cite{ttfcdw} and triggers a series of transitions into
a low-temperature CDW phase.
While some materials such as the Bechgaard salts \cite{tm}, or 
the $TTF$-band of $TTF-TCNQ$ (which has strong $4k_F$-CDW 
fluctuations \cite{ttfcdw}) may well fall into the Luttinger liquid universality
class, it is particularly surprising that CDW systems 
such as $K_{0.3} Mo O_3$, or the $TCNQ$-band in $TTF-TCNQ$, 
should behave as Luttinger liquids. In fact, the photoemission properties
are in striking contrast to the established picture of a fluctuating Peierls
insulator which has been applied quite universally to describe the 
normal state of CDW systems \cite{lra}. It predicts a strongly 
temperature dependent, narrow
[$| \omega | \leq \Delta_{CDW}(T=0)$] pseudogap and  $\rho(q < 0,\omega)$ is
governed by a broadened quasi-particle peak at $\omega < 0$ 
and a weak shadow at $\omega > 0$ \cite{ucla,nic}. 

A Luttinger liquid interpretation for the CDW photoemission
is highly suggestive but  encounters problems which are all resolved in
a Luther-Emery framework. (i) As has been explained before, Luttinger 
liquids have no dominant $2k_F$-CDW correlations: for repulsive interactions
($K_{\rho} < 1$), spin density waves are logarithmically stronger than
CDWs \cite{myrev}, and the behavior of lattice models is consistent with
this picture \cite{hisca}.  
For attractive interactions, the system is dominated by 
superconductivity \cite{myrev}. A spin gap is a
necessary condition for promoting CDW correlations in correlated 1D 
electron systems and is realized in the Luther-Emery model!
(ii) $2k_F$-CDWs often 
are due to electron-phonon coupling, and renormalization group
provides us with a detailed scenario
\cite{myrev,zkl}. The 
dependence of the spin gap on electron-phonon coupling $\lambda$, the
phonon frequency $\omega_D$, and $K_{\rho}$, can be calculated 
reliably \cite{zkl}.
A spin gap also opens if $2k_F$-CDWs are caused by Coulomb interaction between
chains \cite{iccoul}.
(iii) The spin susceptibility of CDW systems above the Peierls 
temperature decreases significantly with
decreasing temperature indicative of activated spin fluctuations.
This applies applies both to $K_{0.3} Mo O_3$ at temperatures from $T_P$
to beyond 700 K \cite{johnston}, and
to the $TCNQ$-chain in $TTF-TCNQ$ where the magnetic
susceptibility contributions of both chains can be separated by NMR
\cite{toshi}. Notice in this context that at finite temperature, the 
density of states in the spin channel of the Luther-Emery model is essentially
the same as for the Lee-Rice-Anderson theory of 
a fluctuating Peierls insulator \cite{evko}, implying that
both models will have similar $\chi(T)$. The temperature-dependent 
susceptibility alone therefore cannot discriminate between these two theories.
Remarkably however, 
in $K_{0.3} Mo O_3$ the conductivity is metallic in the same
temperature range: early experiments over a restricted temperature range
find the resistance $\rho(T) \sim T$ \cite{schnee} while very recent
data taken to much higher temperatures even suggest a sublinear temperature
dependence \cite{bruett} -- not unlike the one found in Luttinger liquids
with repulsive electron-electron interactions \cite{giam}. In $TTF-TCNQ$,
$\rho(T) \sim T$ has been found \cite{cooper}, but it is not known how 
the individual chains contribute to this dependence.
The experiments are  incompatible with the temperature dependence
of the conductivity expected in a fluctuating Peierls insulator \cite{ucla}
which indeed is observed in some organic materials and also $(Ta Se_4)_2 I$.
(iv) For a Luttinger model, the stronger divergence in $\rho(q,\omega)$
is associated with the charge mode and disperses more quickly than the 
weaker signal. In the experiment on $K_{0.3} Mo O_3$, the quickly
dispersing signal is less peaked than the slow one. On the other hand,
the important feature of the Luther-Emery spectral function, Fig.\ 1, 
is that the spin gap supresses the divergence of the charge signal which
disperses more quickly than the divergent spin contribution. 
(v) A CDW transition out of a Luther-Emery liquid by opening a charge
gap at the Peierls temperature, is also consistent
with subtle transfers of spectral weight in regions \em away \rm from
the Fermi energy, observed in spectra taken through the true CDW transition
\cite{dartemp}. In these experiments, the spectral weight at the Fermi 
energy is essentially zero at any temperature. However, at some finite 
energy below $E_F$, the weight drops with a temperature dependence 
consistent with a BCS-like gap. In a naive 
charge-spin separating, Luther-Emery scenario, one would
postulate the opening of a charge gap $\Delta_{\rho}$
at the Peierls temperature
(as a consequence of the establishment of 3D coherence, allowing for the
finite-$T$ transition), in addition to the preexisting spin gap. Thus one
expects a drop of spectral weight at the Peierls transition in an energy
range between $E_F - \Delta_{\sigma}$ and $E_F - \Delta_{\sigma} - 
\Delta_{\rho}$ which, on a sufficiently coarse temperature scale,
would amount to a shift of the leading edge by $\Delta_{\rho}$. 
More likely, the establishment of 3D coherence will 
destroy to some extent the ideal spin-charge separation of the 1D 
Luther-Emery model, and produce a single CDW gap $\Delta_{CDW} >
\Delta_{\sigma}$ below the transition, both for charges and spins. 

On a quantitative level, there is one major problem for the description
of the normal state
of most CDW systems: the spin gap derived from an analysis of the 
magnetic susceptibilities is much smaller than the spin gaps derived
from the peak maxima of the ARPES signals. At present it is not clear
if this indicates a fundamental problem with a Luther-Emery model (the
problem would however not be solved with any competing theory), if this
is due to some ununderstood effect in the photoemission process, or if
it is due to some extrinsic sample property. In another language, it
is not clear what mechanism is responsible for apparent gaps which 
systematically are a sizable fraction of the valence band widths. 

This phenomenology is not consistent with many other theories proposed
for 1D fermions. Theories based on a fluctuating Peierls insulator
would have to explain the two dispersing bands seen in $K_{0.3} Mo O_3$ 
as two separate bands. Two such bands indeed exist but the implication would
be that band structure calculations get one of them too narrow 
by a factor of 5, but get the correct dispersion for the other one
\cite{whangbo}. Moreover, they cannot reconcile the activated susceptibility
with the essentially metallic conductivity above the Peierls temperature.

Standard Luttinger liquids \cite{myrev,zacher}, but also the anomalous ground
states obtained from coupling Luttinger chains so long as their low-energy
fixed point is a Fermi liquid \cite{iccoul}, do neither produce 
the CDW correlations, nor the activated susceptibility. Notice, however,
that both transversely coupled Luttinger liquids (Kopietz \em et al. 
\cite{iccoul} \rm) and the 1D Hubbard model \cite{zacher} can, under some
circumstances, produce spectral functions where the peak on the spinon
dispersion is stronger than that on the charge dispersion. They, however,
would predict a Fermi surface crossing of the photoemission signal which
is not observed experimentally, in addition to the problems listed above. 
In the experiments, instead, the dispersing spectral features bend back
from the Fermi energy as $k$ is increased beyond $k_F$, in a manner
strongly reminiscent of the shadow bands discussed before.

Depsite (important) quantitative problems, 
the Luther-Emery spectral function is consistent with the 
photoemission experiments on $K_{0.3} Mo O_3$ and $TTF-TCNQ$, and 
beyond that, the
model is consistent with much of the other experimental phenomenology
available. I emphasize that while the 
agreement of the Luther-Emery spectral function with the observed photoemission
lineshapes certainly is an argument in favor of this model, it is the 
consistency of its predictions with most other experiments available which 
suggests that it might be a natural starting point for a
description of the low-energy physics of these CDW materials. 

Obviously, this suggestion is somewhat speculative and independent support
is called for. Its virtue is that it comes
to grips with the puzzle that the spin susceptibilities of $K_{0.3} Mo O_3$
and $TTF-TCNQ$
decrease with decreasing temperature while the conductivity are metallic, 
that it leaves space for the good description of optical
properties as a  fluctuating Peierls insulator (they only probe the charge
fluctuations which will form CDW precursors at temperatures much below the
spin gap opening, presumably as a consequence
of emerging 3D coherence), and that it 
provides an (admittedly phenomenological)
description of the photoemission properties of these materials with extremely
1D \em electronic \rm properties \cite{pouget}. 
As in the Bechgaard salts \cite{tm}, 
a single-particle exponent 
$\alpha \sim 1/2 \ldots 1$ would be required implying strong long-range
electron-electron interactions, and there is at best preliminary
support from transport measurements \cite{bruett}, for
such strong correlations in  $K_{0.3} Mo O_3$. 
Retarded electron-phonon coupling
could increase $\alpha$ over its purely electronic value \cite{zkl}.
To what extent this mechanism contributes could be gauged from the measured
$\alpha$ which must be larger than the one derived from the enhancement of
$v_{\rho}$ over the band velocity (h\'{e}las strongly depending on the
accuracy of band structure calculations).
In $TTF-TCNQ$, the analysis is made difficult by the presence of two
chains. There is evidence for strong long-range electron-electron
interactions on the $TTF$-chain from the observation of $4k_F$-CDW
fluctuations, but the situation for $TCNQ$ is less clear. If a sizable
enhancement of the dispersion of the ARPES signals over the estimated
bandwidths can be interpreted as evidence for long-range electronic
correlations, they would indeed be present on both chains.

\section{Summary}
In this paper, I have presented a construction of the dynamical 
correlation functions of the 1D Luther-Emery model. This model has
one gapped degree of freedom, and an ungapped one, and describes
1D superconductors and Peierls insulators (spin gap) and 1D Mott
insulators (charge gap). It is a natural extension of Luttinger
liquid theory to the peculiar
phase intermediate between metal and band insulator, made possible
in one dimension by the phenomenon of charge-spin separation.
The dynamical correlation functions presented here show where and to what
extent the two typically 1D features of a Luttinger liquid: charge-spin
separation, and anomalous dimensions of operators, survive in the
presence of a gap in one channel. Since an exact calculation of such
correlation functions usually is not possible in a Luther-Emery model,
our construction relied heavily on limiting cases, symmetries, and
equivalences to other models. However, it successfully passed several
tests in situations where exact results were available from other
methods.

The main emphasis of the paper was on the single-particle spectral function
which is measured in photoemission. We showed that, generically, 
charge-spin separation and anomalous dimensions are also visible in the
spectral functions of the Luther-Emery model. Specifically, for a spin
gapped system with repulsive interactions, describing a 1D charge density
wave system, the spectral function has a true singularity on the gapped
spin dispersion with an anomalous exponent $\alpha - 1/2$ while on the
charge dispersion, the Luttinger liquid divergence is cut off to a finite
maximum by the spin gap -- a results which finds a straightforward 
explanation in terms of convolution of charge and spin correlation functions.
For attractive interactions, i.e. a 1D superconductor, two divergences
with anomalous dimensions are found. For 1D Mott insulators, i.e. a charge
gap, one finds one or two inverse-square-root singularities, i.e. no
anomalous dimension (due to spin-rotation invariance), depending on
the order of the velocities of the charge or spin fluctuations. It was
also shown how these procedures can be generalized to two- and multi-particle
correlation functions. 

Besides predicting spectral functions for the many 1D models falling into
the Luther-Emery universality class, 
there are a few experimental situations where these results can be usefully
applied. They successfully describe the photoemission spectrum of the
1D Mott insulator $Sr Cu O_2$ \cite{kim}, to an extent leaving few questions
open, the most notable one being experimental resolution. Less clearcut
but perhaps more interesting are CDW materials such as $K_{0.3} Mo O_3$
and $TTF-TCNQ$ which show very unusual photoemission spectra. 
These are qualitatively consistent with a Luther-Emery model, and we have
proposed that these materials might, most naturally, be described in
this framework. A Luther-Emery phase is necessary as an intermediate
between a Luttinger liquid and a long-range ordered CDW, and 
$K_{0.3} Mo O_3$ and $TTF-TCNQ$ are natural candidates for searching for such
a strange metal. This scenario requires strong electron-electron 
interactions at least at high energies, and not all CDW materials 
need fall into this 
scheme. If the electron-phonon interaction is so strong as to produce
CDW precursor fluctuations at very high temperature, and the electronic
correlations are weak enough, the establishment
of a Luttinger liquid, and the crossover to a Luther-Emery liquid 
at lower temperature, may be quenched, and a fluctuating Peierls insulator 
\cite{lra} or a bipolaron liquid \cite{pascal}
may be a more appropriate picture. Some CDW materials such
as $(Ta Se_4)_2 I$ \cite{tase}, (perylene)$_2 PF_6$ \cite{ucla}, 
(fluoranthene)$_2
PF_6$ \cite{beijing} apparently are consistent with this picture.
However, $K_{0.3} Mo O_3$ and $TTF-TCNQ$ are not consistent, and the
consistency of the spectral functions constructed in this paper
with the published experiments, and the
analysis of further experiments indicate that, besides electron-phonon
coupling, electronic correlations must be important in these CDW systems.

\acknowledgements
I wish to acknowledge fruitful discussions with J. W. Allen, 
M. Grioni, M. Gul\'{a}csi, G.-H. Gweon, D. Malterre, and J.-P.
Pouget. I am supported by  DFG under SFB 279-B4 and as a Heisenberg fellow.

\figure{FIG.~1 Dispersion of peaks in the spectral function $\rho(q,\omega)$
of a spin-gapped Luther-Emery model with $v_{\rho} > v_{\sigma}$. The dispersion
laws $\varepsilon_{\rho}(q)$ and $\varepsilon_{\sigma}(q)$ are given in the
text. The heavy solid and dashed lines give the signals in the main band
[sign($\omega$) = sign($q$)] while the light dashed lines label the shadow
bands [sign($\omega$) = -- sign($q$)]. } 

\figure{FIG.~2 Spectral function of the spin-gapped Luther-Emery model for $q<0$.
$v_{\rho} > v_{\sigma}$ has been assumed, as applies to a 1D Peierls insulator.
The dashed line at $-\varepsilon_{\rho}(k)$ indicates the Luttinger liquid
divergence which is suppressed here to a finite maximum.}

\figure{FIG.~3 Dispersion of singularities in the spectral function 
$\rho(q,\omega)$
of a spin-gapped Luther-Emery model with $v_{\rho} < v_{\sigma}$. The solid 
lines give the signals in the main band while the dashed lines label the shadow
bands. } 

\figure{FIG.~4 Spectral function $\rho(q,\omega)$
of the spin-gapped Luther-Emery model 
with $v_{\rho} < v_{\sigma}$ (applying to 1D superconductors) for $q<0$.}

\figure{FIG.~5 Spectral function $\rho(q,\omega)$ 
of the charge-gapped Luther-Emery model, describing 1D Mott insulators, 
with $v_{\rho} > v_{\sigma}$ for $q<0$. The dispersions of the signals follow
Fig.~3 with 
$\rho \leftrightarrow \sigma$ everywhere, and the straight dashed
lines in the shadow bands must be ignored. }


\begin{references}
\bibitem{early} When the gap becomes small in a physical situation, 
	the high-energy tail of these particle-hole excitations becomes 
	activated again. It is unclear, however, if this effect is correctly 
	described by using the earlier Ansatz \cite{brief1}
	with $K_{\sigma} \neq 1$.
\end{references}
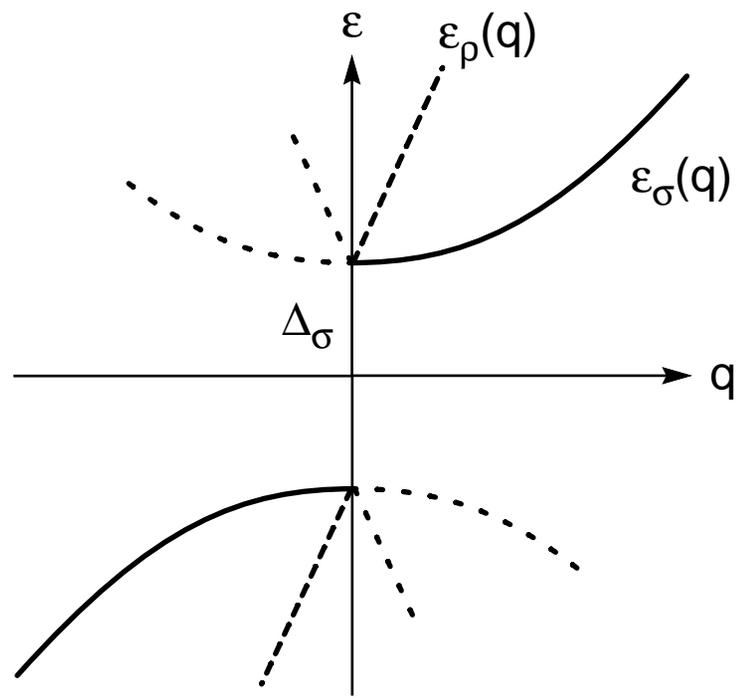
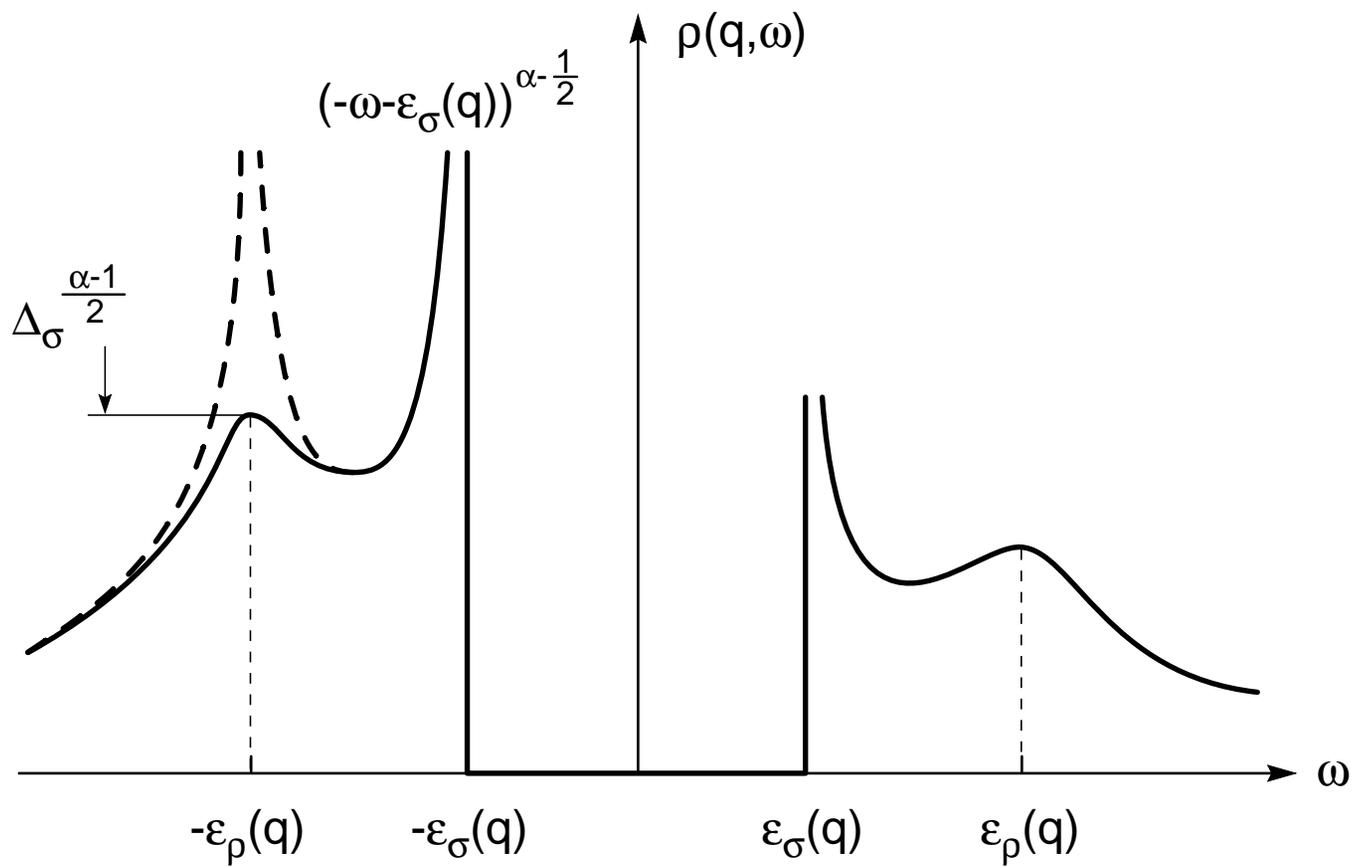
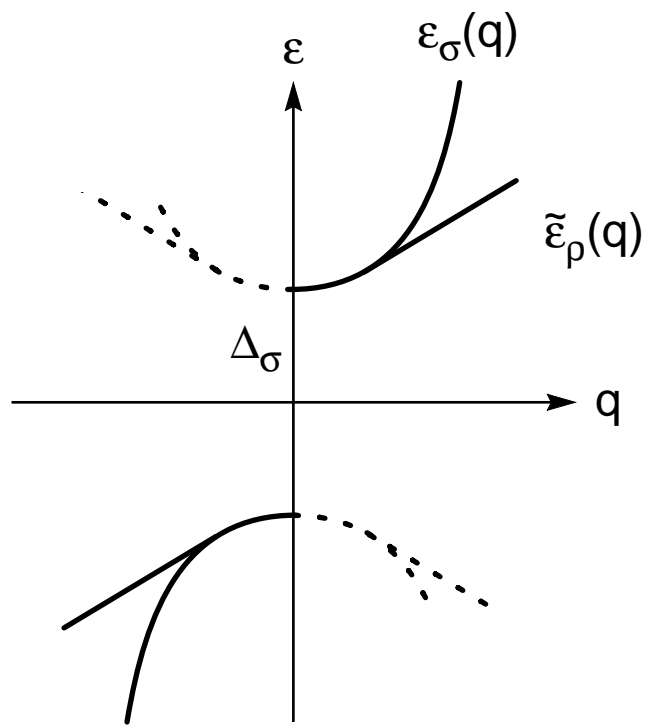
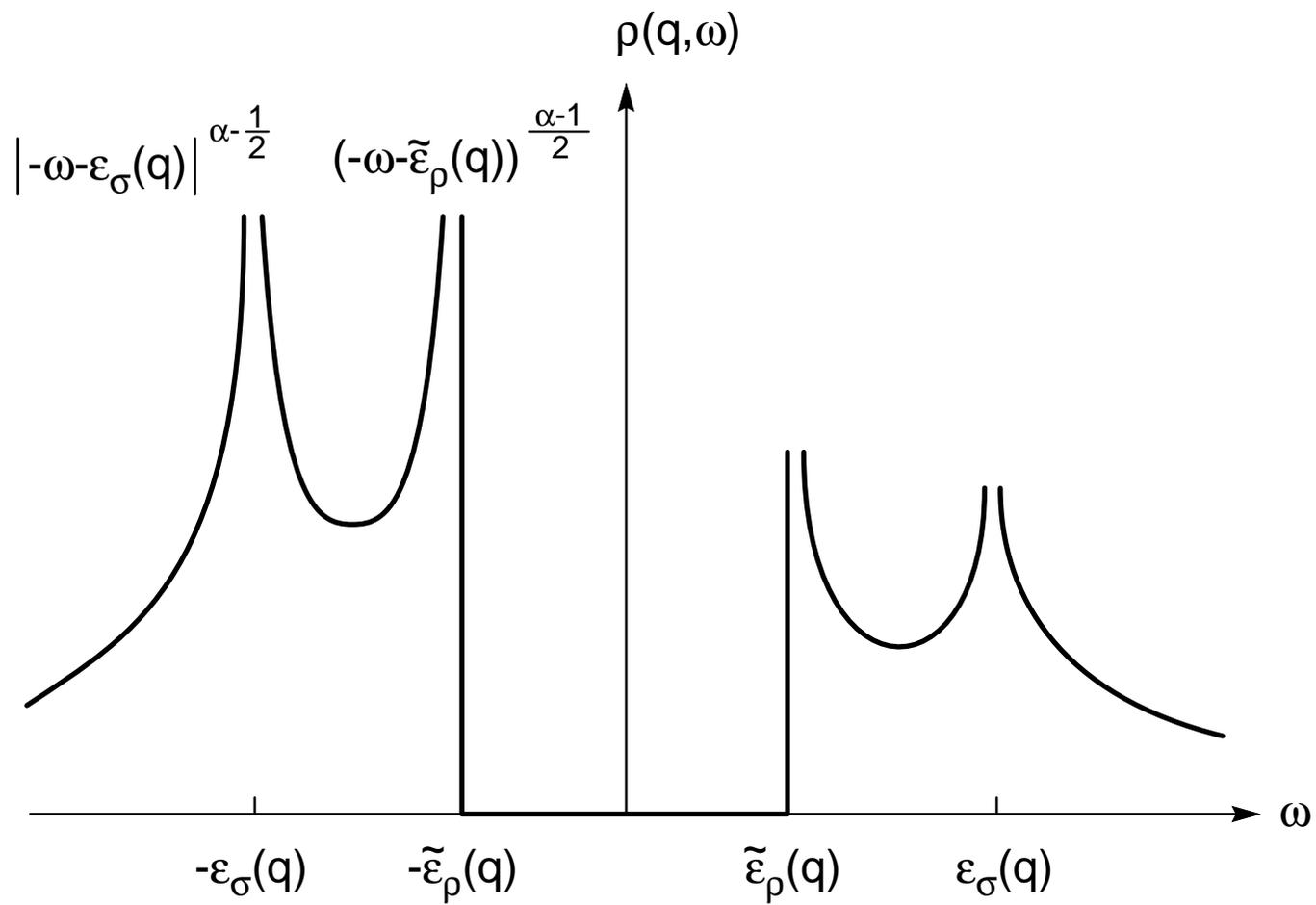
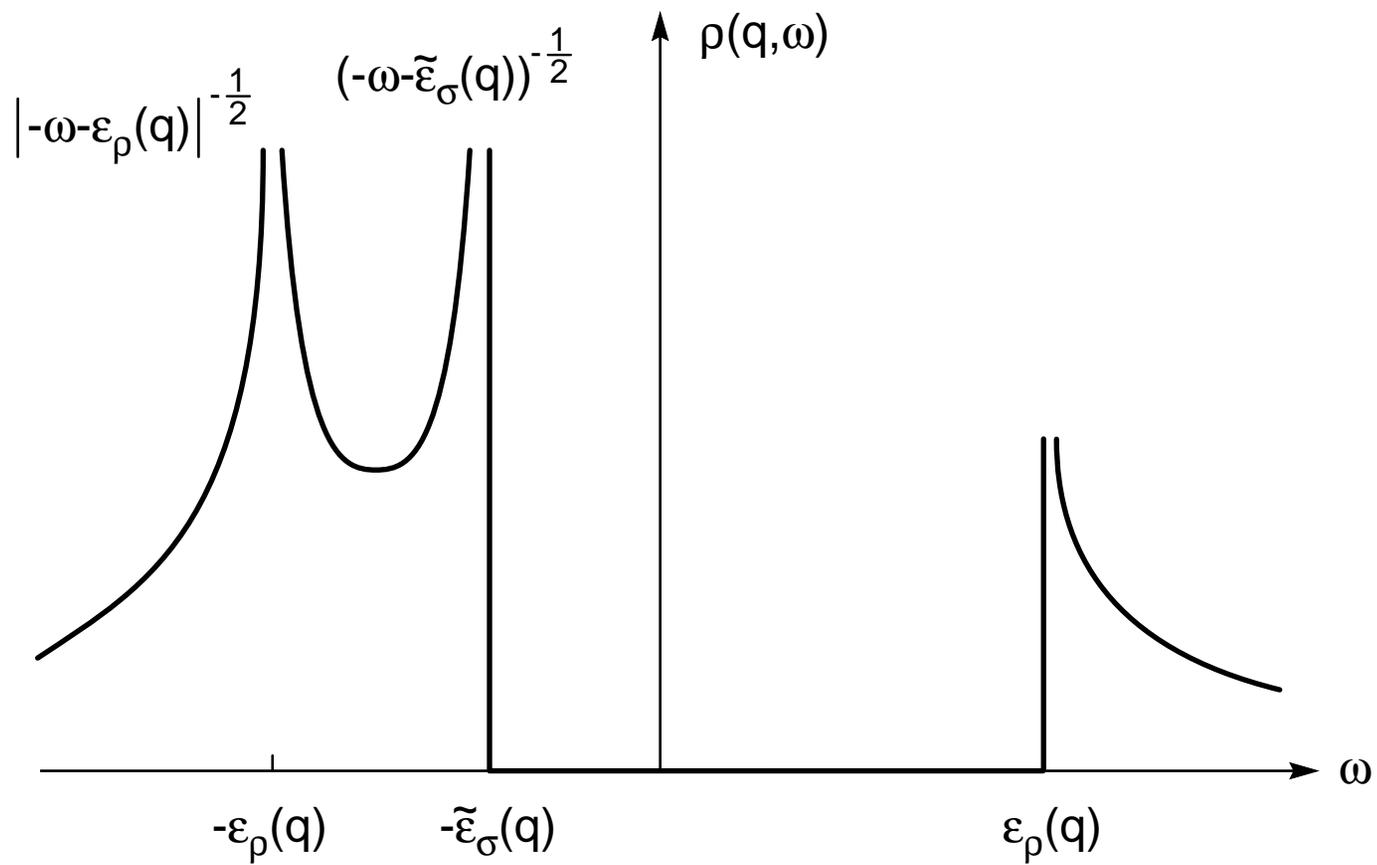
\end{document}